\RequirePackage{fix-cm}
\documentclass[smallextended]{svjour3}       
\smartqed  
\usepackage{graphicx}
\usepackage{latexsym}
\usepackage{amssymb}
\begin{document}

\title{Times of arrival (TOA) of signals in the Kerr-MOG black hole}


\author{G.Y. Tuleganova$^{1}$ \and R.N. Izmailov$^{1}$ \and R.Kh. Karimov$^{1}$  \and A.A.Potapov$^{2}$ \and K.K. Nandi$^{1, 2, 3}$}


\institute{G.Y. Tuleganova \at
              \email{gulira.tuleganova@yandex.ru}
           \and
           R.N. Izmailov \at
              \email{izmailov.ramil@gmail.com}
           \and
           R.Kh. Karimov \at
              \email{karimov\_ramis\_92@mail.ru}
           \and
           A.A.Potapov \at
              \email{a.a.potapov@strbsu.ru}
           \and
           K.K. Nandi \at
              \email{kamalnandi1952@yahoo.co.in}
           \and
           $^{1}$ Zel'dovich International Center for Astrophysics, M.Akmullah Bashkir State Pedagogical University, 3A, October Revolution Street, Ufa 450008, RB, Russia
           \at
           $^{2}$ Department of Physics \& Astronomy, Bashkir State University, 47A, Lenin Street, Sterlitamak 453103, RB, Russia
           \at
           $^{3}$ High Energy Cosmic Ray Research Center, University of North Bengal, Darjeeling 734 013, WB, India
}

\date{Received: date \today / Accepted: date}

\maketitle

\begin{abstract}
Modified gravity (MOG) theories are alternatives to general relativity (GR) that arose primarily from the need to explain the observed galactic flat rotation curves without invoking the elusive dark matter hypothesized by GR. A well known MOG is the Scalar-Tensor-Vector-Gravity (STVG) developed by Moffat, who has also found a spinning solution called the Kerr-MOG black hole (BH) characterized by the spin $a$ and MOG parameter $\alpha$, the latter determining the strength of the gravitational vector forces. We consider the static-MOG metric ($a=0$) to first understand how the nature of geometry drastically changes depending on different sectors of $\alpha $. Then we study the influence of $\alpha $ in each sector on a new astrophysical diagnostic caused by \textit{frame dragging}, viz., the difference $\Delta t$ in the times of arrival (TOA) at the observer of signals emanating from a variable pulsar (PSR) passing behind a Kerr-MOG lens in a PSR-BH binary system. The study generalizes the zeroth order Laguna-Wolszczan formula up to third PPN order in $\left(1/r\right)$ using thin-lens approximation, which reveals how $\Delta t$ is influenced both by $a$ and $\alpha$. The magnitude and sign of $\alpha $ indicate deviations from GR ($\alpha =0$) and future measurements may constrain $\alpha$ provided a suitable binary is identified.
\end{abstract}

\section{Introduction}
\label{intro}
It is well known that the explanations, of the observed galactic flat rotation curves (for the earliest observation on rotation curves, see \cite{Oort:1931}) and of the 1998 discovery of cosmological acceleration \cite{Riess:1998,Perlmutter:1998}, within general relativity (GR) require the postulate of attractive dark matter and repulsive dark energy respectively, both invisible. However, despite several speculations (WIMPs, scalar fields, etc) about the particle content of dark matter, all experiments to date have failed to directly detect any of them (see, e.g., \cite{Agnese:2014}). Therefore, physicists have been led to devise alternative theories, collectively called modified gravity (MOG) theories, that preserve the successes of GR but do not require the postulate of elusive dark matter. Such MOG includes, but not limited to, Weyl conformal gravity \cite{Mannheim:2011} (see also \cite{Nandi:2012}), MOdified Newtonian Dynamics (MOND) \cite{Milgrom:1983} and $f(R)$-gravity \cite{Nojiri:2006}. Along this line, a well known MOG is the Scalar-Tensor-Vector-Gravity (STVG), developed by Moffat \cite{Moffat:2006}, that is ideologically distinct from GR, and has led to far-reaching consequences as evidenced by numerous works till date \cite{Izmailov:2019a,Brownstein:2006a,Brownstein:2006b,Moffat:2013a,Perez:2017,Lopez:2017,Brownstein:2007,Moffat:2008,Moffat:2013b,Moffat:2015a,Moffat:2014,Moffat:2015b,Moffat:2015c}. Several merits of the STVG are worth mentioning: It possesses a stable vacuum and the Hamiltonian is bounded from below. Even though the action is not gauge invariant, it can be shown that the theory is free of ghosts.

The spinning solution of STVG, also proposed by Moffat \cite{Moffat:2015b}, is called the Kerr-MOG black hole (BH) \cite{Brownstein:2006a}. The metric is characterized by spin $a$ and a massive vector field with an enhanced Newtonian acceleration defined by a gravitational constant $G=G_{N}(1+\alpha )$, where $G_{N}$ is the Newtonian gravitational constant, $\alpha $ is the MOG parameter representing a Yukawa-like force that is repulsive for $\alpha >0$. With the identification $q=\sqrt{\alpha G_{N}}M$, where $M$ is the mass of the BH, the Kerr-MOG solution formally resembles the Kerr-Newman or in the static case, the Reissner-Nordstr\"{o}m BH of GR. However, the \textit{physical content} of the Kerr-MOG metric is quite different from that of GR as the quantity $q$ is not the independently prescribed Coulomb charge of GR. On the contrary, it can be called a gravitational charge $q$ determined solely by the mass itself. The effect of $q$ is symbolized by a gravitational Yukawa-like term with two parameters $\widetilde{\mu }$ and $\alpha$ in the radial acceleration law $a(r)$ as will be given later.

The Kerr-MOG and its static counterpart have been investigated for several effects \cite{Izmailov:2019a,Brownstein:2006a,Brownstein:2006b,Moffat:2013a,Perez:2017,Lopez:2017,Brownstein:2007,Moffat:2008,Moffat:2013b,Moffat:2015a,Moffat:2014,Moffat:2015b,Moffat:2015c} including observable lensing properties \cite{Izmailov:2019a}. Another \textit{new} observable that we study here is the times of arrival (TOA) of
signals under thin-lens approximation, denoted by $\Delta t$, caused by the frame dragging of the spinning lens. The TOA, variantly called relative time delay, is completely different from the well known Shapiro gravitational time delay\footnote{In contrast, there is also an effect of gravitational \textit{time advancement}, proposed originally in \cite{Bhadra:2010}, and some of its effects have been explored in \cite{Deng:2017,Ghosh:2015}.}, in which two way\ elapsed times are added and averaged, whereas in the TOA, the light travel times are subtracted. The stage to look for the TOA is set by the recent speculation of different binary lens systems in which we assume the Kerr-MOG BH to be the spinning lens partner. The concepts, of TOA and of thin lens approximation, to be used in this paper are briefly elaborated below.

TOA is a potential observable that samples the frame dragging caused by a spinning lens. Consider a binary system, where a variable light source $S$ orbits a spinning compact object (BH lens). Suppose two light rays emerge from the source $S$ behind a spinning lens $L$ (with mass $M$, spin $a$), pass on either side of it to reach the observer at $O$, say, the Earth. The rays will reach at different times at $O$ and this difference in the times of arrival (TOA) is caused by the frame dragging due to the intervening spinning lens. The dragging causes the light path lengths on either side of the lens to differ, shorter on the co-rotating side and longer on the counter-rotating side (see Fig.1). The TOA\ loosely resembles the astrophysical analogue of the quantum Bohm-Aharonov effect\footnote{The analogy is loose because the light rays do not travel in vacuum but in the weak gravitational field. For a true gravitational Bohm-Aharonov effect, see \cite{Ford:1981}.}.

The TOA of pulses has first been calculated to the zeroth order by Laguna and Wolszczan \cite{Laguna:1997} in the Kerr metric for some hypothetical binary systems. A similar, though not exactly the same, type of effect was studied by Datta and Kapoor \cite{Datta:1985}, where light rays were assumed to emerge not from a variable source behind the lens but from two diametrically opposite points on a spinning compact astrophysical object itself. A good example of TOA could be the early observation of extremely rapid fluctuations in the brightness of quasar $1525+227$ with characteristic time scale $\sim $ $200$ sec speculated to be caused by a spinning black hole of mass $M\sim 5\times 10^{8}M_{\odot }$ situated between the quasar and the observer \cite{Matilsky:1982}. Recently, TOA has been theoretically studied for the Johannsen metric \cite{Izmailov:2019b} as a possible observable diagnostic to test the validity of the so-called "no-hair" conjecture of Penrose \cite{Penrose:1969}. The effect of string parameter on TOA was investigated in \cite{Izmailov:2020}. So far no precise experimental data on TOA are available but very accurate future data from suitably identified binaries can in principle constrain the values of MOG parameter $\alpha$. Such constraints were recently discussed also in the context of other MOG theories \cite{Karimov:2018}.

Thin-lens approximation (see the details, e.g., in Hartle \cite{Hartle:2003}) applies to a situation where the source, lens and observer are all considered as points and the light rays are assumed to travel in straight lines with the deflection taking place only at the lens. This works excellently when the rays travel vast distances compared to the lens size with the impact parameter far larger than the photon sphere. Hence the relevant angles making the quadrilateral in Fig.1 with the optical axis $SLO$ are small. A more accurate calculation of TOA should of course involve integration between two finite points on the optical axis of the exact null geodesics around the spinning lens. Nonetheless, the thin-lens approximation provides a simple and elegant description of physical reality without significantly compromising rigor \cite{Hartle:2003}.

In this paper, we shall theoretically study the effect of spin $a$ and the Kerr-MOG BH parameter $\alpha $ on TOA in binary lens systems assuming the source, lens and observer to be aligned. Specifically, we want to see how $\Delta t$ differs in magnitude from the Kerr value ($\alpha =0$) of GR for a given lens, when its mass $M$ and spin $a$ are independently specified. The new results that follow include (i) the\ re-interpretation of the static MOG metric as a brane-world BH with $\alpha $ (or $q_{0}$) playing the role of the brane-world tidal charge, (ii)\ the generalization of the zeroth order Laguna-Wolszczan law for $\Delta t$, to \textit{finite} distance scales on the optical axis that are actually realizable in experiments. Numerical values will be tabulated for two speculated binary lens systems. The advantage of the method adopted in this paper is that it can be applied with ease to other spinning metrics available in the literature (see, e.g., \cite{Manko:1992,Glampedakis:2006}). We shall take $c=1$ unless specifically restored.

In Sec.2, we shall revisit the STVG theory with its assumptions and the Kerr-MOG BH. In Sec.3, we briefly point out the nature of the static-MOG metric for different ranges of $\alpha$ including its connection to the brane-world. In Sec.4, we derive the equations for the TOA. In Sec.5, we shall integrate the straight null path trajectories between two fixed points on the optical axis using the thin-lens approximation. Sec.6 presents indicative numerical estimates for two binary lens systems and Sec.7 concludes the paper.

\section{The STVG theory and Kerr-MOG BH}
\label{sec:1}
We shall briefly revisit the STVG theory \cite{Moffat:2006} for explaining the involved terms and rewrite the Kerr-MOG BH for our use. The action in STVG theory is \cite{Moffat:2013a}
\begin{equation}
S=S_{\mathrm{GR}}+S_{\phi }+S_{\mathrm{S}}+S_{\mathrm{M}},
\end{equation}%
where
\begin{eqnarray}
S_{\mathrm{GR}} &=& \frac{1}{16\pi}\int d^{4}x\sqrt{-g}\frac{1}{G}R, \\
S_{\phi} &=& -\int d^{4}x\sqrt{-g}\left(\frac{1}{4}B^{\mu\nu}B_{\mu\nu} - \frac{1}{2}\widetilde{\mu}^{2}\phi^{\mu}\phi_{\mu}\right), \\
S_{\mathrm{S}} &=& \int d^{4}x\sqrt{-g}\frac{1}{G^{3}}\left[\frac{1}{2}g^{\mu\nu}\nabla_{\mu}G\nabla_{\nu}G - V(G)\right] \nonumber\\
&& +\int d^{4}x\frac{1}{\widetilde{\mu}^{2}G}\left[\frac{1}{2}g^{\mu\nu}\nabla_{\mu}\widetilde{\mu}\nabla_{\nu}\widetilde{\mu} -V(\widetilde{\mu})\right].
\end{eqnarray}
Here, $g_{\mu\nu}$ is the spacetime metric, $R$ denotes the Ricci scalar, and $\nabla_{\mu}$ is the covariant derivative, $\phi^{\mu}$ stands for a Proca-type massive vector field,
\begin{equation}
B_{\mu\nu} = \partial_{\mu}\phi_{\nu} - \partial_{\nu}\phi_{\mu},
\end{equation}%
$G(x)$ and $\widetilde{\mu}(x)$ are dynamical scalar fields that vary in space and time, $V(G)$, and $V(\widetilde{\mu})$ are the corresponding potentials. The term $S_{\mathrm{S}}$ is the massless scalar field action and $S_{\mathrm{M}}$ refers to possible matter sources, which we set to zero. The full energy-momentum tensor is
\begin{equation}
T_{\mu\nu}=T_{\mu\nu}^{\mathrm{M}}+T_{\mu\nu}^{\phi }+T_{\mu\nu}^{\mathrm{S}},
\end{equation}%
where
\begin{eqnarray}
T_{\mu\nu}^{\mathrm{M}} &=& -\frac{2}{\sqrt{-g}}\frac{\delta S_{\mathrm{M}}}{\delta g^{\mu\nu}}, \\
T_{\mu\nu}^{\phi} &=& -\frac{2}{\sqrt{-g}}\frac{\delta S_{\phi}}{\delta g^{\mu\nu}}, \\
T_{\mu\nu}^{\mathrm{S}} &=& -\frac{2}{\sqrt{-g}}\frac{\delta S_{\mathrm{S}}}{\delta g^{\mu\nu}}.
\end{eqnarray}
The Kerr-MOG\ BH were found under the following assumptions \cite{Izmailov:2019a}: (i) The mass of the vector field $\phi^{\mu}$ is neglected; since the effects of $m_{\phi}$ manifest only at kiloparsec scales from the source \cite{Moffat:2014}, so it can be disregarded when solving the field equations for compact objects such as a BH. (ii) $G$ is a constant that depends on the parameter $\alpha$ \cite{Moffat:2006}:
\begin{equation}
G = G_{N} \left(1 + \alpha\right),
\end{equation}%
where $G_{N}$ denotes Newton's gravitational constant, and $\alpha$ is a free adimensional MOG parameter.

Given these assumptions, the action (1) takes the simplified form
\begin{equation}
S = \int d^{4}x\sqrt{-g}\left( \frac{R}{16\pi G}-\frac{1}{4}B^{\mu\nu}B_{\mu\nu}\right).
\end{equation}%
Variation of the action with respect to $g_{\mu v}$ now yields the STVG field equations:
\begin{equation}
G_{\mu\nu} = -8\pi GT_{\mu\nu}^{\phi},
\end{equation}%
where $G_{\mu\nu}$ is the Einstein tensor, and the energy-momentum tensor for the vector field $\phi^{\mu}$ is given by \cite{Moffat:2006}
\begin{equation}
T_{\mu\nu}^{\phi} = \frac{1}{4}\left(B_{\mu}{}^{\alpha }B_{\nu\alpha }-\frac{1}{4}g_{\mu\nu}B^{\alpha\beta}B_{\alpha\beta}\right).
\end{equation}%
Varying the action (11) with respect to the vector field $\phi^{\mu}$, we obtain the dynamical equation for such field as
\begin{equation}
\triangledown_{\nu}B^{\mu\nu}=0.
\end{equation}%
The equation of motion for a test particle in coordinates $x^{\mu}$ is given by
\begin{equation}
\frac{d^{2}x^{\mu}}{d\tau^{2}} + \Gamma_{\alpha\beta}^{\mu}\frac{dx^{\alpha}}{d\tau}\frac{dx^{\beta}}{d\tau} 
= \frac{q}{m}B_{\nu}^{\mu}\frac{dx^{\nu}}{d\tau },
\end{equation}%
where $\tau$ denotes the particle proper time, $m$ is the test particle mass and $q$ is the coupling constant with the vector field.

Moffat \cite{Moffat:2015b} postulates that the gravitational charge $q$ of the vector field $\phi^{\mu}$ is proportional to the BH mass $M$, i.e.,
\begin{equation}
q=\pm \sqrt{\alpha G_{N}}m.
\end{equation}%
Particles fall freely and satisfy the equivalence principle but do not follow geodesic motion. We see that in STVG theory the nature of the gravitational field has thus been modified with respect to GR in two ways: a change in the gravitational constant $G = G_{N}\left(1+\alpha\right)$, and a vector field $\phi^{\mu}$ that exerts a gravitational Lorentz-type force on any material object through Eq.(15).

The line element of the Kerr-MOG BH of mass $M$ and spin angular momentum $J=aM$ is \cite{Perez:2017}:
\begin{equation}
d\tau ^{2}=g_{tt}dt^{2}+2g_{t\varphi }dtd\phi -g_{rr}dr^{2}-g_{\theta\theta}d\theta ^{2}-g_{\phi \phi }d\varphi ^{2},
\end{equation}
where
\begin{eqnarray}
g_{tt} &=&\frac{c^{2}(\Delta -a^{2}\sin ^{2}\theta )}{\rho ^{2}}, \\
g_{t\varphi } &=&\frac{ac\sin ^{2}\theta }{\rho ^{2}}\left[\left(r^{2}+a^{2}\right) -\Delta \right] , \\
g_{rr} &=&\frac{\rho ^{2}}{\Delta }, \\
g_{\theta \theta } &=&\rho ^{2}, \\
g_{\phi \phi } &=&\left[ \left( r^{2}+a^{2}\right) ^{2}-\Delta a^{2}\sin^{2}\theta \right] \frac{\sin ^{2}\theta }{\rho ^{2}}, \\
\Delta &=&r^{2}-\frac{2GMr}{c^{2}}+a^{2}+\frac{\alpha G_{N}GM^{2}}{c^{4}}=r^{2}-\frac{2G_{N}(1+\alpha )Mr}{c^{2}}+a^{2}+\frac{\alpha (1+\alpha
)G_{N}^{2}M^{2}}{c^{4}},  \nonumber \\
\rho ^{2} &=&r^{2}+a^{2}\cos ^{2}\theta .
\end{eqnarray}%
Kerr BH of GR is recovered at $\alpha =0$.

\section{Static MOG BH}
\label{sec:2}
The purpose of this section is the following: We shall be taking a range of values of $\alpha$ for the computation of TOA below and from the ensuing analysis, we shall now know which sector of $\alpha$ represents what type of geometry and what are the differences, if any, in the values of TOA. According to Moffat's postulate \cite{Moffat:2015b} the positive value ($\alpha >0$) produces a repulsive gravitational Yukawa-like force and in the simplest case of spherical symmetry yields a first order weak field solution $\phi_{0}$ for the $t-$component of the vector field $\phi_{\mu}$ (see for details \cite{Perez:2017}):
\begin{equation}
\phi_{0} = -\alpha G_{N}M\left( \frac{\exp (-\widetilde{\mu }r)}{\widetilde{\mu }r}\right) ,
\end{equation}%
where $\widetilde{\mu}$ is an independent mass parameter. With this field, the Eq.(15) yields the radial acceleration%
\begin{equation}
a(r)=-\frac{G_{N}(1+\alpha )M}{r^{2}}+G_{N}M\alpha \frac{\exp (-\widetilde{\mu }r)}{r^{2}}(1+\widetilde{\mu }r).
\end{equation}%
The Yukawa-type contribution is evident in the last term. Newtonian gravity is recovered in the weak field limit $\widetilde{\mu }r<<1$. Then the static MOG BH is given by \cite{Moffat:2006}
\begin{eqnarray}
ds^{2} &=&-A(r)c^{2}dt^{2}+B(r)dr^{2}+C(r)\left( d\theta ^{2}+\sin^{2}\theta d\varphi ^{2}\right) ,  \nonumber \\
A(r) &=&\frac{1}{B(r)}=1-\frac{2G_{N}(1+\alpha )M}{c^{2}r}+\frac{G_{N}^{2}M^{2}(1+\alpha )\alpha }{c^{4}r^{2}}, \\
C(r) &=&r^{2}.
\end{eqnarray}%
Redefining the mass $M$ in relativistic units and a dimensionless charge $Q_{0}$ as
\begin{eqnarray}
m_{\bullet } &=&\frac{GM}{c^{2}}=\frac{G_{N}(1+\alpha )M}{c^{2}}, \\
Q_{0} &=&\frac{\alpha }{1+\alpha},
\end{eqnarray}%
we can rewrite the metric (25-27) in the form
\begin{eqnarray}
A(r) &=&1-\frac{2m_{\bullet }}{r}+\frac{Q_{0}m_{\bullet }^{2}}{r^{2}}, \\
C(r) &=&r^{2}.
\end{eqnarray}%
Scaling the above metric further by $r_{\mathrm{s}}=2m_{\bullet }=\frac{2G_{N}(1+\alpha )M}{c^{2}}=1$, we write
\begin{eqnarray}
A(r) &=&\frac{1}{B(r)}=1-\frac{1}{r}+\frac{Q}{r^{2}}, \\
C(r) &=&r^{2}, \quad Q=\frac{1}{4}Q_{0}=\frac{1}{4}\left( \frac{\alpha }{1+\alpha }\right).
\end{eqnarray}%
The spacetime metric form (32,33) resembles a \textit{brane-world} BH for $-1<\alpha <0$ or $Q<0$ obtained by Dadhich et al. \cite{Dadhich:2000} belonging to the brane-world theory developed by Sasaki et al. \cite{Sasaki:2000}. To see the resemblance more clearly, note that the effective field equations on the brane is given by%
\begin{equation}
R_{\mu \nu }=-\mathcal{E}_{\mu \nu },\quad R_{\mu }^{\mu }=0=\mathcal{E}_{\mu }^{\mu },\quad \nabla ^{\mu }\mathcal{E}_{\mu \nu }=0
\end{equation}%
where the symmetric $\mathcal{E}_{\mu\nu}$ is the limit on the brane of the projected bulk Weyl tensor. Dadhich et al. \cite{Dadhich:2000} solution explicitly is
\begin{eqnarray}
A(r) &=&\frac{1}{B(r)}=1-\left( \frac{2M}{M_{p}^{2}}\right) \frac{1}{r}+\left( \frac{q_{0}}{\widetilde{M}_{p}^{2}}\right) \frac{1}{r^{2}}, \\
C(r) &=&r^{2}, \quad q_{0}=Q\widetilde{M}_{p}^{2},
\end{eqnarray}%
where $q_{0}$ is the \textit{tidal charge} imprint from the bulk onto the brane, $M_{p}$ and $\widetilde{M}_{p}$ are the Planck masses on the brane and the 5-d bulk respectively and typically, $\widetilde{M}_{p}$ $<<M_{p}$. In the short distance limit (small $r$), perturbative analysis, which starts from an exactly anti-de Sitter background, shows that \cite{Dadhich:2000}%
\begin{equation}
q_{0}=-\frac{M}{\widetilde{M}_{p}}
\end{equation}%
so that the tidal charge is negative. It can be easily seen that, when $q_{0}<0$, there is only one horizon with an area larger than its Schwarzschild counterpart, which implies that bulk effects enhance the entropy and reduce the temperature of the black hole.

Following the Keeton and Petters method \cite{Keeton:2006} of PPN expansion for the metric (30,31), we can obtain the light deflection for the impact parameter $b$ (details omitted)
\begin{eqnarray}
\delta \varphi  &=&A_{1}\frac{m_{\bullet }}{b}+A_{2}\left( \frac{m_{\bullet}}{b}\right) ^{2}+A_{3}\left( \frac{m_{\bullet }}{b}\right) ^{3}+... \\
A_{1} &=&4,\quad A_{2}=(5-Q_{0})\left( \frac{3\pi }{4}\right) , \quad%
A_{3}=\frac{128}{3}-16Q_{0}.
\end{eqnarray}%
The geometry changes drastically depending on the sectors of $\alpha $ as follows: The condition for a \textit{naked singularity} is $A_{2}<3\pi$ \cite{Keeton:2006}, which yields $Q_{0}>1$ that in turn corresponds to the range $-\infty <\alpha <-1$. The metric (32,33) has \textit{no central} \textit{singularity} if $\alpha >0$, or $Q_{0}>0$ because of Yukawa-type gravitational repulsion on test particles near $r=0$. The value $\alpha =-1$ or $Q_{0}=\infty $ is excluded since it produces $m_{\bullet }=0$ or a trivial flat spacetime. The quantity $Q_{0}<0$ or $-1<\alpha <0$ giving the brane-world BH has been interpreted as negative tidal charge $q_{0}<0$ arising out of the free gravitational field propagating in the bulk. The effect of the tidal charge on the galactic halo has been studied in the literature \cite{Nandi:2009}. The static-MOG and the brane-world BH parameters are joined at the GR divide at $\alpha =0$, hence the BHs are in a sense complementary to each other though they belong to entirely different parent theories. This is a remarkable feature of static MOG\ BH and we assume that the above distinct geometrical characters will be preserved even under spin of the source.
\begin{figure}[!ht]
  \includegraphics[scale=0.8]{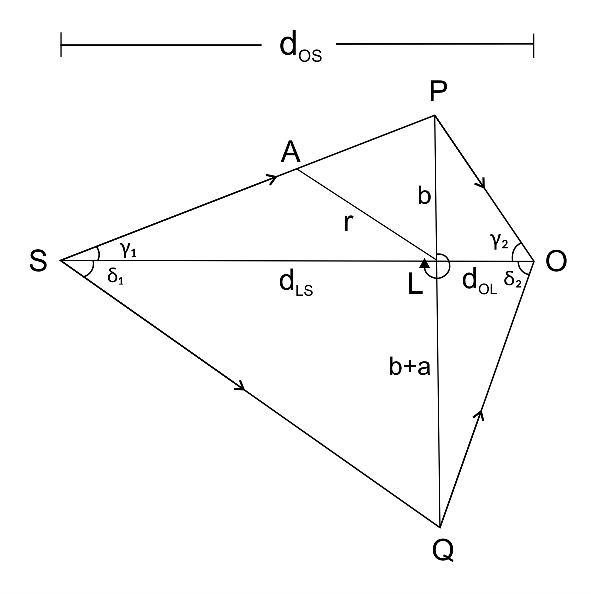}
\caption{The generic thin-lens slender quadrilateral (here angles are exaggerated). $S$, $L$ and $O$ are the source, lens and observer respectively aligned on a straight line ($\beta =0$), $b$ is the impact parameter and $a$ is the spin. The arbitrary angles are as shown.}
\label{fig:1}
\end{figure}
The generic thin-lens slender quadrilateral (here angles are exaggerated). $S$, $L$ and $O$ are the source, lens and observer respectively aligned on a straight line ($\beta =0$), $b$ is the impact parameter and $a$ is the spin. The arbitrary angles are as shown.

\section{TOA in Kerr-MOG BH}
\label{sec:3}
To derive the equation for TOA, let us return to the generic spinning metric (17). We assume that light rays are emanating from behind the spinning BH. Thus, the null trajectory on the equatorial plane ($\theta =\pi /2$) given by $d\tau ^{2}=0$, so that the coordinate time required for light rays along an infinitesimal null world line is given by
\begin{equation}
dt_{\pm }=\frac{d\phi }{g_{tt}}[-g_{t\phi }\pm h(r,\phi )],
\end{equation}%
where
\begin{equation}
h(r,\phi )\equiv \sqrt{g_{t\phi }^{2}-g_{tt}\left\{ g_{rr}\left( \frac{dr}{%
d\phi }\right) ^{2}+g_{\phi \phi }^{2}\right\} }.
\end{equation}%
We assume the passage of coordinate time to be positive for both $\pm$ sides of the lens identifying $d\phi >0$, for light rays passing the lens by the co-rotating side ($+$) and $d\phi <0$ for the counter-rotating side ($-$), so that $dt_{+}$ and $dt_{-}$ are both positive. The net difference between the two null rays in the time of arrival (TOA) at the observer is given by:
\begin{equation}
\left\vert dt\right\vert =dt_{-}-dt_{+}=\frac{\left\vert d\phi \right\vert }{%
g_{tt}}[g_{t\phi }+h(r,\phi )]-\frac{\left\vert d\phi \right\vert }{g_{tt}}%
[-g_{t\phi }+h(r,\phi )]=\frac{2g_{t\phi }}{g_{tt}}\left\vert d\phi
\right\vert .
\end{equation}%
This delay $dt$ is due to the frame-dragging effect characterized by$\left(\frac{2g_{t\phi}}{g_{tt}}\right)$, which we are going to compute in this
paper. We assume that the source, spinning lens and the observer are aligned, that is, they are situated on a straight line (see Fig.1). When the lens is not spinning, the path lengths of the light ray on both sides of the lens would be the same and there would be no difference in TOA at the observer. However, when the lens is spinning the path lengths will differ - longer for co-rotating and shorter for counter-rotating rays - giving rise to the TOA. The rays on the equatorial plane pass are required to pass through the weak field so that the thin-lens approximation applies, that is the light rays pass on either side of the spinning lens at large $r$ ensuring, for a given lens of mass $M$ and spin $a$, that the quantity $\left( \frac{M}{r}\right) <<1$.

We shall apply the generic formula (42) to Kerr-MOG BH (18-23) by expanding $\left(\frac{2g_{t\phi}}{g_{tt}}\right)$ such that, to first order in $a$
and up to third PPN order in $\left(M/r\right) $, it yields:
\begin{equation}
\left\vert dt\right\vert = |d\phi|\left(\frac{1}{c}\right)\left[\frac{4aM}{r}\psi_{1} + \frac{2aM^{2}}{r^{2}}\psi_{2}
+ \frac{8aM^{3}}{r^{3}}\psi_{3}\right],
\end{equation}%
where
\begin{equation}
\psi _{1}=(1+\alpha ),\quad \psi _{2}=(1+\alpha )(4+3\alpha ),\quad \psi_{3}=(1+\alpha )^{2}(2+\alpha)
\end{equation}%
are the parameters showing deviations away from Kerr BH having fixed values $%
\psi _{1}=1$, $\psi _{2}=4$, $\psi _{3}=2$ corresponding to $\alpha =0$. The
effect of $\alpha $ is evident from Eqs.(43,44). For a given mass $M$ and
spin $a$, the measured data on $dt$ would in principle constrain $\alpha $.

The total TOA $\Delta t$ between two null rays traveling from the source
(PSR) to observer along two opposite sides of the intermediate spinning lens
(BH) is%
\begin{eqnarray}
\Delta t &=&\left( \frac{1}{c}\right) \int\limits_{0}^{\pi }d\phi \left[
\frac{4aM}{r}\psi _{1}+\frac{2aM^{2}}{r^{2}}\psi _{2}+\frac{8aM^{3}}{r^{3}}%
\psi _{3}\right]  \nonumber \\
&\equiv &\frac{1}{c}\left( I_{1}+I_{2}+I_{3}\right) =\Delta t_{1}+\Delta
t_{2}+\Delta t_{2}.
\end{eqnarray}%
We compute the integral locating the spinning lens at the origin of a polar
system of coordinates on the equatorial plane ($\theta =\pi /2$). In the
following, we shall derive explicit expressions for $\Delta t_{1}$, $\Delta
t_{2}$ and $\Delta t_{3}$ within the finite distance thin-lens approximation.

\section{Thin-lens approximation}
\label{sec:4}
The approximation means that the light bending takes place at a radial
distance much smaller than the typical distances $d_{OL}$, $d_{OS}$, $d_{LS}$
such that light rays can be assumed to propagate in straight lines when far
away from the lens, with the bending taking place only at the point lens
\cite{Hartle:2003}. In the present Kerr-MOG BH situation, the approximation requires three
additional non-trivial conditions as enumerated below.

(i) Light rays emerging from $S$, after passing along line segments on
either side of spinning lens $L$ in the equatorial plane, should meet
exactly at $O$ making a quadrilateral $SPOQ$ of Fig.1. We show that this is
possible only under the assumption that $(a/b)<<1$ such that orders of $%
(a/b)^{2}$ and higher can be neglected. Thin-lens approximation implies
that, for a given $d_{OL}$, the impact parameter $b$ and the ratio $\chi $ ($%
=\frac{d_{LS}}{d_{OL}}$) be such that the angles $\gamma _{1},\gamma
_{2},\delta _{1}$,$\delta _{2}$ $<<1.$Following the Boyer-Lindquist formula
\cite{Boyer:1967} for the bending of light at $P$ and $Q$ respectively, we have%
\begin{eqnarray}
\theta _{1} &=&\frac{4M}{b}\left( 1-\frac{a}{b}\right) ,\mathrm{ (prograde
motion of light)} \\
\theta _{2} &=&\frac{4M}{b}\left( 1+\frac{a}{b+x}\right) ,\mathrm{ (retrograde
motion of light)}
\end{eqnarray}%
where $x(a,b)$ is an unknown function to be determined. The formation of the
quadrilateral $SPOQ$ of Fig.1 is possible only if the following condition is
satisfied by $x(a,b)$:%
\begin{equation}
-bx(x+b)+a(b^{2}+bx+x^{2})=0.
\end{equation}%
Solving the equation under the condition that $x(a,b)=0$ at $a=0$ and
neglecting $\left( \frac{a}{b}\right) ^{2}$ and higher orders, we get a root
$x_{1}=a$, yielding the impact parameters $b$ and $b+a$ as in Fig.1. The
same result has been used, without proof, for the Kerr BH by Dymnikova \cite{Dymnikova:1986}.

(ii) In order to be in the weak field regime, light rays should pass far
away from the photon spheres $r_{\mathrm{ph}}^{\pm }$ appearing respectively
on the co-rotating ($+$) and counter-rotating sides ($-$) of the lens. These
radii $r_{\mathrm{ph}}^{\pm }$ demarcate natural strong field limits perceived
by null rays. Note that the photon spheres around BHs, the deflection angles
become logarithmically divergent, as a result of which the light rays get
eternally captured there \cite{Bozza:2002}. Since light rays are passing far away from
the location of the photon sphere, a strictly exact expression for radius of
$r_{\mathrm{ph}}\ $is not needed for our purpose \cite{Sasaki:2000}. Still, to have an idea
of how far is far, we can estimate the Kerr-BH values of the radii of the
photon sphere $r_{\mathrm{ph}}^{\pm \mathrm{Kerr}}$ for a given $M$ and $a$:%
\begin{equation}
r_{\mathrm{ph}}^{\pm \mathrm{Kerr}}=2M\left[ 1+\cos \left\{ \frac{2}{3}\arccos
\left( \frac{\mp a}{M}\right) \right\} \right] .
\end{equation}

(iii) To avoid the breakdown of the thin lens approximation, we assume that
the smaller of the two impact parameters, $b$ and $b+a$, must far exceed the
larger of the two radii $r_{\mathrm{ph}}^{\pm \mathrm{Kerr}}$, so, let us take $%
b=10^{n}r_{\mathrm{ph}}^{-}$, where $n>1$ is any real number. The value of $b$
should be so chosen as to the preserve the smallness of the involved angles
in Fig.1.

In Fig.1, we have by construction $d_{LS}=\chi d_{OL}$, where $\chi >0$ is a
finite constant ratio between the two distances, $PLQ\perp OLS$, and the
arbitrary angles are as indicated. The radial distance is measured from the
lens $L$. After a bit of algebra involving piecewise integration of typical
straight null trajectories in the relevant sectors, corotating $\left( \frac{%
1}{r_{\mathrm{cor}}}\right) $ and counterrotating $\left( \frac{1}{r_{\mathrm{cou%
}}}\right) $, we get the final result by subtracting the integrals over the
path lengths, viz., $SQO-SPO$:%
\begin{equation}
I_{1}=4aM\left[ \int\limits_{\pi /2}^{\pi }\frac{1}{r_{\mathrm{cor}}}d\phi
+\int\limits_{0}^{\pi /2}\frac{1}{r_{\mathrm{cor}}}d\phi \right] -4aM\left[
\int\limits_{-\pi /2}^{-\pi }\frac{1}{r_{\mathrm{cou}}}d\phi
+\int\limits_{0}^{-\pi /2}\frac{1}{r_{\mathrm{cou}}}d\phi \right] .
\end{equation}%
The integrals $I_{1}$, $I_{2}$ and $I_{3}$ of Eq.(45) yield%
\begin{eqnarray}
\Delta t_{1} &=&\frac{I_{1}}{c}=\frac{8aM}{cb\chi d_{OL}F}\left\{ 2a\left[
b\left( \chi -1\right) +\chi d_{OL}\right] +b\left[ b\left( \chi -1\right)
+2\chi d_{OL}\right] \right\} \psi _{1}, \\
\Delta t_{2} &=&\frac{I_{2}}{c}=\frac{aM^{2}}{cd_{OL}^{2}\chi ^{2}B}\left[
B\pi -\chi A+\left\{ A+\left[ B+\left( a^{2}+2ab+2b^{2}\right) d_{OL}^{2}%
\right] \pi \right\} \chi ^{2}\right] \psi _{2}, \\
\Delta t_{3} &=&\frac{I_{3}}{c}=\frac{8aM^{3}}{cd_{OL}^{2}}\left[ 4+\frac{%
4d_{OL}^{3}}{b^{3}}+\frac{4d_{OL}^{3}}{F^{3}}-\frac{4}{\chi ^{3}}+\frac{C}{%
b^{2}\chi }+\frac{C}{F^{2}\chi }+\frac{D}{b\chi ^{2}}+\frac{D}{F\chi ^{2}}%
\right] \psi _{3}
\end{eqnarray}%
where%
\begin{eqnarray}
A &=&2bd_{OL}\left( a+b\right) \left( a+2b\right) , \\
B &=&b^{2}\left( a+b\right) ^{2}, \\
C &=&3d_{OL}^{2}\left( \chi -1\right) , \\
D &=&3d_{OL}\left( 1+\chi ^{2}\right) , \\
F &=&a+b.
\end{eqnarray}

The Eqs. (51-58) are the generic equations for the delay. Eq.(51) is the
desired generalization of the original Laguna-Wolszczan formula \cite{Laguna:1997}:
\begin{equation}
\Delta t_{1}\simeq \frac{16aM}{cb},
\end{equation}%
which is obtained at $\chi =1$, $\alpha =0$. We shall now plug the lens
parameter values $a,M$, distance values $b$, $d_{OL}$, $\chi =d_{LS}/d_{OL}$
into the Eqs. (51-58), take care to preserve the smallness of the angles $%
b/d_{OL}$, $b/d_{LS}<<1$, and tabulate the TOA components $\Delta t_{1},$ $%
\Delta t_{2}$ and $\Delta t_{3}$, all of them containing the the MOG
parameter $\alpha $ via $\psi _{1},\psi _{2}$ and $\psi _{3}$ (Tables 1,2).

\section{Numerical estimates}
\label{sec:5}
Pulsar-BH binary systems provide the best laboratory for testing the TOA
predictions since variable sources like pulsars, quasars, GRBs etc can give
out signals at the instant they are behind the spinning black hole on the
optical axis $OL$ and their times of arrival $\Delta t$ can be measured at
the observer. Though a concrete example of such a binary is yet to be
detected, the prospects for discovery seem quite promising \cite{Laguna:1997}. We assume
the pulsar orbit to be in the equatorial plane of the Kerr-MOG BH and the
line of sight to be perpendicular to the axis of the spin.

\subsection{PSR-Cygnus X-1 binary}
\label{sec:5a}

Of all pulsars discovered so far, an early estimate was that a small but
significant number of them belong to a PSR-BH category with a BH having
masses a few times of solar masses \cite{Lipunov:2005}. We consider a typical illustration,
namely, of a PSR-Cygnus X-1 binary with $M=14.8M_{\odot }=2.19\times 10^{6}$
cm, $a=0.95M=2.08\times 10^{6}$ cm, $d_{OL}=1.86$ kpc $=5.74\times 10^{21}$
cm \cite{Gou:2011}. The Kerr BH case corresponds to $\alpha =0,$ $\psi _{1}=1$, $\psi
_{2}=4$, $\psi _{3}=2$. The values from Eq.(49), viz., $r_{\mathrm{ph}%
}^{-}=8.66\times 10^{6}$ cm and $r_{\mathrm{ph}}^{+}=3.03\times 10^{6}$ cm,
shows that $r_{\mathrm{ph}}^{-}$ is the larger of the two radii. Accordingly,
to preserve the thin-lens and PPN approximation, we choose $b=10^{4}r_{\mathrm{%
ph}}^{-}=8.66\times 10^{10}$ cm, so that $\frac{M}{b}$ $\sim 10^{-5}$
justifying the PPN expansion. The lens-source distances $d_{LS}=\chi d_{OL}$
in Fig.1 can be varied by varying $\chi $ but preserving the required
smallness of the angles (in radian): $\gamma _{1}\simeq b/d_{LS},\delta
_{1}\simeq $ $(a+b)/d_{LS}$, $\gamma _{2}\simeq b/d_{OL},\delta _{2}\simeq $
$(a+b)/d_{OL}$. In this paper, to fix values, we choose a certain $\chi =0.1$
since the values of $\Delta t_{i}$ are not very sensitive to values of $\chi
$.

\begin{table}[!ht]
\begin{tabular}{|c|c|c|c|}
\hline
$\alpha $ &  $\Delta t_{1}$ ($\mu $sec) & $\Delta t_{2}$ ($\mu $sec) & $\Delta t_{3}$ ($\mu $sec) \\ \hline
$-3$ &  $-5.64\times 10^{-2}$ & $2.80\times 10^{-6}$ & $-9.62\times 10^{-11}$ \\ \hline
$-2.5$ & $-4.23\times 10^{-2}$ & $1.47\times 10^{-6}$ & $-2.70\times 10^{-11}$ \\ \hline
$-2$ &  $-2.82\times 10^{-2}$ & $0.56\times 10^{-6}$ & $0$ \\ \hline
$-1.5$ & $-1.41\times 10^{-2}$ & $0.07\times 10^{-6}$ & $0.30\times 10^{-11}$ \\ \hline
$-1$ & $0$ & $0$ & $0$ \\ \hline
$-0.5$ & $1.41\times 10^{-2}$ & $0.35\times 10^{-6}$ & $0.90\times 10^{-11}$ \\ \hline
$0$ & $2.82\times 10^{-2}$ & $1.12\times 10^{-6}$ & $4.81\times 10^{-11}$ \\ \hline
$0.5$ & $4.23\times 10^{-2}$ & $2.31\times 10^{-6}$ & $13.54\times 10^{-11}$ \\ \hline
$1$ & $5.64\times 10^{-2}$ & $3.92\times 10^{-6}$ & $28.88\times 10^{-11}$ \\ \hline
$1.5$ & $7.06\times 10^{-2}$ & $5.95\times 10^{-6}$ & $52.66\times 10^{-11}$ \\ \hline
$2$ & $8.47\times 10^{-2}$ & $8.41\times 10^{-6}$ & $86.66\times 10^{-11}$ \\ \hline
$2.5$ & $9.88\times 10^{-2}$ & $11.28\times 10^{-6}$ & $132.70\times 10^{-11}$ \\ \hline
$3$ & $11.29\times 10^{-2}$ & $14.58\times 10^{-6}$ & $192.58\times 10^{-11}$ \\ \hline
\end{tabular}
\caption{The table shows some typical values of TOA components $\Delta t_{1}$, $\Delta t_{2}$, $\Delta t_{3}$ for the PSR-Cygnus X-1 binary. The last three columns contain the effect of the MOG parameter $\alpha$ through $\psi_1$, $\psi_2$ and $\psi_3$ as in Eqs. (51), (52) and (53).}
\end{table}

Table 1 shows that, to leading order, the TOA component $\Delta t_{1}$ would
be at the $\mu $sec level that could be measurable in the future. The other
components $\Delta t_{2}$ and $\Delta t_{3}$ however hold no promise to be
measurable even in the far future. The Kerr values appear are at $\alpha =0$%
. The values of $\alpha $ are chosen taking into account the ranges of $%
\alpha $ discussed in Sec.3.

The third column shows, measurement of $\Delta t_{1}$ could be possible in
the near future though still technically quite challenging. However, as is
evident from the last two columns, unless the accuracy of measurement of
total observed delay is raised to pico-second level and higher, which is
absurd today, there is little hope to measure the components $\Delta t_{2}$
and $\Delta t_{3}$.

\subsection{PSR-SgrA* binary}
\label{sec:5b}

Recent observations suggest that there are probably $\sim 100$ pulsars surrounding the supermassive spinning BH SgrA* with orbital periods $\lesssim $ $10$ years \cite{Pfahl:2004} and a few among them are expected to form PSR-BH binaries with stellar sized BH companions residing within $\sim 1$ parsec of SgrA* \cite{Faucher:2011}. We shall assume the possibility that some of the pulsars cross the optical axis $OLS$ behind SgrA* making a PSR-SgrA* binary.

\begin{table}[!ht]
\begin{tabular}{|c|c|c|c|}
\hline
$\alpha $ & $\Delta t_{1}$ ($\mu $sec) & $\Delta t_{2}$ ($\mu $sec) & $\Delta t_{3}$ ($\mu $sec) \\ \hline
$-3$ & $-8.42$ & $4.75\times 10^{-7}$ & $-1.85\times 10^{-14}$ \\ \hline
$-2.5$ & $-6.31$ & $2.49\times 10^{-7}$ & $-0.52\times 10^{-14}$ \\ \hline
$-2$ & $-4.21$ & $0.95\times 10^{-7}$ & $0$ \\ \hline
$-1.5$ & $-2.10$ & $0.11\times 10^{-7}$ & $0.05\times 10^{-14}$ \\ \hline
$-1$ & $0$ & $0$ & $0$ \\ \hline
$-0.5$ & $2.10$ & $0.59\times 10^{-7}$ & $0.17\times 10^{-14}$ \\ \hline
$0$ & $4.21$ & $1.90\times 10^{-7}$ & $0.92\times 10^{-14}$ \\ \hline
$0.5$ & $6.31$ & $3.92\times 10^{-7}$ & $2.61\times 10^{-14}$ \\ \hline
$1$ & $8.42$ & $6.66\times 10^{-7}$ & $5.57\times 10^{-14}$ \\ \hline
$1.5$ & $10.52$ & $10.11\times 10^{-7}$ & $10.16\times 10^{-14}$ \\ \hline
$2$ & $12.63$ & $14.27\times 10^{-7}$ & $16.73\times 10^{-14}$ \\ \hline
$2.5$ & $14.74$ & $19.15\times 10^{-7}$ & $25.62\times 10^{-14}$ \\ \hline
$3$ & $16.84$ & $24.74\times 10^{-7}$ & $37.18\times 10^{-14}$ \\ \hline
\end{tabular}
\caption{The table shows some typical values of TOA components $\Delta t_{1}$, $\Delta t_{2}$, $\Delta t_{3}$ for the PSR-SgrA* binary. The last three columns contain the effect of the MOG parameter $\alpha$ through $\psi_1$, $\psi_2$ and $\psi_3$ as in Eqs. (51), (52) and (53).}
\end{table}

Table 2 shows some typical TOA components $\Delta t_{1},\Delta t_{2},\Delta t_{3}$ calculated from Eqs.(51-58) for SgrA* with values given by Kato \textit{et al.} \cite{Kato:2010}, viz., $M=4.2\times 10^{6}M_{\odot }$, $d_{OL}=7.6$ kpc and a \textit{unique} $a=0.44M$ so that $r_{\mathrm{ph}}^{-}=2.15\times 10^{12}$ cm, $r_{\mathrm{ph}}^{+}=1.51\times 10^{12}$ cm, $b=10^{7}r_{\mathrm{ph}}^{-}$ so that $M/b<<1$. The distances $d_{LS}$, $d_{OL}$ in Fig.1 are such that the angles remain small: $\gamma _{1}\simeq $ tan$\gamma _{1}=b/d_{LS}\simeq \delta _{1}\simeq $ tan$\delta _{1}$ etc. The Kerr values appear at $\alpha=0$. The values of $\alpha $ are chosen taking into account its ranges discussed in Sec.3 and taking $\chi =0.1$.

Thus, for the PSR-SgrA* binary, we find that $\Delta t_{1}$ $\sim$ a few $\mu$sec, which should be measurable with current technology provided an appropriate PSR orbiting SgrA* is identified in the future missions.

\section{Conclusions}
\label{concl}
Assuming the Kerr-MOG BH to be a possible spinning lens partner in a PSR-BH binary system, we investigated the effect of the MOG parameter $\alpha$ on the new diagnostic of TOA using gravitational thin-lens approximation and PPN expansion up to third order in $\left( 1/r\right) $. We clarify that the TOA samples frame dragging and is fundamentally different from the Shapiro time delay in which the two-way light travel times are added, whereas in the TOA the travel times of two light rays from behind the spinning lens are subtracted. To calculate TOA in the Kerr-MOG, we took into account some important factors that are as follows:

First, from the closed slender quadrilateral in Fig.1, it follows that to first order in $\left(\frac{a}{b}\right)$, the impact parameters must be $b$ and $b+a$ on the relevant sides of the lens, which automatically takes into account the frame dragging effect via spin $a$. Second, all distances are taken to be finite as required by any realistic astrophysical lensing experiment. The lens-source distance ($d_{LS}$) has been taken ten times closer to the lens than the vast observer-lens distance ($d_{OL}$) using $d_{LS}=\chi d_{OL}$, with $\chi =0.1$. This choice is motivated by practical considerations, where the orbiting pulsar is expected to orbit the lens pretty closer than the distance $d_{OL}$. Thirdly, the null rays must pass at large distance away from the photon spheres $r_{\mathrm{ph}}^{\pm}$ on either side of the lens in order to avoid capture at the photon sphere radii on either side. So we considered $b=10^{n}r_{\mathrm{ph}}^{-}$, where $n>1$ is a large number and $r_{\mathrm{ph}}^{-}$ is always the larger of the two radii $r_{\mathrm{ph}}^{\pm}$ such that $\frac{M}{b}<<1$, as required.

The order of magnitude estimates for $\Delta t_{1}$, $\Delta t_{2}$ and $\Delta t_{3}$ from Eqs.(51-58) are tabulated for two illustrative binary lens systems. They are quite robust, that is, the order of magnitudes remain unaltered even when the distance ratios $\chi $ is varied preserving small angles required by the thin-lens approximation. The TOA estimates are presented in Tables 1 and 2, are only indicative since no binary system has yet been conclusively identified, although\ recent surveys suggest that the number of PSR-BH binaries could be significant \cite{Ford:1981}. We have considered two possible binary systems, PSR-Cygnus X1 and PSR-SgrA* and values of $\Delta t_{1}$ have been found to be at the $\mu $sec level. The expression for our $\Delta t_{1}$ Kerr-MOG generalizes the Laguna-Wolszczan formula (51) for Kerr BH. However, the higher order delays $\Delta t_{2}$ and $\Delta t_{3}$ are too tiny to be measurable.

Remarkably, it follows that despite two distinct regimes, viz., repulsive Yukawa regime ($\alpha >0$), the brane-world or attractive Yukawa regime ($-1<\alpha <0$) and the naked singularity regime ($-\infty <\alpha <-1$), the TOA is still of the same order of magnitude, $\Delta t_{1}\sim 10^{-2}$ $\mu$sec (Table 1) for PSR-Cygnus X1, which could be measured in the near future since a precision of $0.1$ $\mu $sec has already been achieved \cite{VanStraten:2001}. We point out that the negative signs in the latter two regimes are of no consequence since we calculated only the relative difference in arrival times. As to the PSR-SgrA* binary, it is found that $\Delta t_{1}$ $\sim$ a few $\mu $sec (Table 2), which should be measurable even today provided a suitable variable source is detected from among the pulsars orbiting SgrA* \cite{Pfahl:2004,Faucher:2011}.

Even though the leading order TOA $\Delta t_{1}$[Eq.(51)] could already mean a potential new test of general relativity, the strong field lens effects involving the logarithmically divergent deflection of light near the photon sphere could reveal surprisingly new characteristics. This is a task we leave for the future.

\begin{acknowledgements}
The reported study was funded by RFBR according to the research Project No.
18-32-00377.
\end{acknowledgements}



\end{document}